\documentclass[aps,prl,twocolumn,showpacs,amsmath,amssymb
,nofootinbib,showkeys,floatfix]{revtex4}

\usepackage{dcolumn}   
\usepackage{bm}        

\usepackage{latexsym}
\usepackage{amsmath}
\usepackage{amssymb}

\usepackage{supertabular} 
\usepackage{placeins}
\usepackage{epsfig}
\usepackage{graphicx}
\usepackage{graphics}

\usepackage{hyperref} 
\hypersetup{
   colorlinks=true,
   linkcolor=blue,
   citecolor=blue,
   urlcolor=blue
}

\begin{document}
\title{The $N/D$ method with non-perturbative left-hand-cut discontinuity and the $^1S_0$ $NN$ partial wave}
\author{D.R. Entem}
\affiliation{Grupo de F\'{\i}sica Nuclear and IUFFyM,
Universidad de Salamanca, E-37008 Salamanca,
Spain
}
\author{J.A. Oller}
\affiliation{Departamento de F\'\i sica,
Universidad de Murcia, E-30071 Murcia,
Spain
}

\begin{abstract}
 In this letter we deduce an integral equation that allows to calculate the exact left-hand-cut discontinuity
 for an uncoupled $S$-wave partial-wave amplitude in potential scattering for a given finite-range potential.  
 The results obtained from the $N/D$ method for the partial-wave amplitude are rigorous, since now 
the discontinuities along the left-hand cut and right-hand cut are exactly known. 
This solves 
 the  open question with respect to the $N/D$ method and
 the effect on the final result of the non-perturbative iterative diagrams in the evaluation of $\Delta(A)$.
A 
 big advantage of the method is that  short-range physics
 (corresponding to integrated out degrees of freedom within low-energy Effective Field Theory) 
does not contribute to $\Delta(A)$ and it manifests through the extra subtractions  
 that are implemented within the  method. 
 We show the equivalence of the $N/D$ method and the Lippmann-Schwinger (LS) equation 
 for a nonsingular $^1S_0$ $NN$ potential (Yukawa potential).
The equivalence between the $N/D$ method with one extra subtraction and 
 the LS equation renormalized with one counter term or with subtractive renormalization
 also holds for the singular attractive higher-order ChPT potentials. 
The $N/D$ method also allows to evaluate partial-wave amplitudes with a higher number of extra subtractions,  
that we fix  in terms of shape parameters within the effective range expansion. The result at NNLO shows
that the $^1S_0$ phase shifts might be accurately described once the electromagnetic
and charge symmetry breaking terms are included. Our present results can be extended to higher partial waves as well as to 
coupled channel scattering.

\end{abstract}

\maketitle


Since Weinberg proposal in the 1990s~\cite{plb.251.288,npb.363.3,plb.295.114} 
to use the framework of Chiral
Effective Field Theory for the nucleon-nucleon ($NN$) system, many works have been published on this topic.
The first attempts were performed by Ord\'o\~nez, Ray and van Kolck~\cite{prl.72.1982,prc.53.2086} giving potentials in
coordinate space. The M\"unich group calculated the perturbative $NN$ amplitude at NNLO~\cite{npa.625.758} and
Epelbaum, Gl\"ockle and Mei{\ss}ner developed an NNLO potential~\cite{npa.637.107,npa.671.295}
that was widely used in nuclear systems. The first high-precision Chiral potential
was developed by Entem and Machleidt at N$^3$LO~\cite{plb.524.93,prc.68.041001}. Recently an N$^4$LO potential has been
developed by Epelbaum {\it et al.}~\cite{prl.115.122301} and even high partial waves has been studied at
N$^5$LO~\cite{prc.92.064001}.

Since the very beginning renormalization issues have been pointed out 
and many works have been published on this topic and on power counting 
issues~\cite{plb.424.390,plb.464.169,npa.677.313,npa.725.85,plb.568.109,ijmpa.21.1079,plb.580.149,prc.72.054002,plb.636.305,
prc.72.054006,prc.74.054001,epja.32.77,prc.83.024003,npa.653.209,prc.80.044002,
prc.85.034002,prc.77.044006,fbs.54.2191,prc.88.054002,arxiv:1606.01489}.
However one of the main controversies 
 is the range in which the regulator cut-off used to regularize the singular interactions should be used.
 One point of view is to use regulators on a higher scale than the low energy scale of the 
 Effective Field Theory (EFT), i.e., the pion mass ($m_\pi$), but lower than the high energy scale, i.e., $\Lambda_\chi\sim 1$ GeV.
 This approach  works very well phenomenologically~\cite{prc.88.054002,prl.115.122301}, although the results 
 depend on the actual value used for the cut-off  and some cut-off artifacts can be seen.
 The other point of view is the one used in the standard renormalization procedure, by 
 taking the cut-off to infinity. Along this line, works based on renormalization with boundary 
conditions~\cite{plb.580.149,prc.72.054002}, subtractive renormalization~\cite{npa.653.209,prc.80.044002} and
renormalization with one counter term have been shown to be
equivalent~\cite{prc.77.044006}, though they are not very successful phenomenologically~\cite{fbs.54.2191}.

Another possibility to obtain regularization independent results is to use the $N/D$
method \cite{chew.119.467}, since no regulators are needed even for singular interactions. The method only
uses the  unitarity and analytical properties of partial wave amplitudes 
 and gives rise to a  linear integral equation (IE), from which the scattering amplitude can be calculated.
  The IE stems from  appropriate dispersion relations, 
which can include extra subtractions. The input for the $N/D$ IE  is the
discontinuity of the partial-wave projected $T$ matrix along the left-hand-cut (LHC), that we denote by $\Delta(A)$ where $A=k^2$ 
(with $k$ the on-shell center of mass three-momentum). 
This discontinuity  stems from the explicit degrees of freedom  included in the theory.
 In this way, an advantage of the method is that the counter terms of the EFT, i.e., zero-range interactions
 that are allowed by symmetry 
 and that absorb the infinities generated by loop diagrams, do not give any
contribution to this discontinuity.
 Nonetheless, their physical effects can be
 reproduced by including  appropriate subtraction constants, 
up to reaching the desired accuracy. 
Related to this, since $\Delta(A)$ is the dynamical input when referring 
to the order of our calculation we strictly refer to the order in which the LHC discontinuity of the potential is evaluated, which 
coincides with the Weinberg counting. 
 The traditional shortcut of the $N/D$ method up to now is that for a given potential 
 the discontinuity $\Delta(A)$ is not known a priori, and the approximation typically made is
 to calculate it perturbatively.
 This approach has been pursued by Oller {\it et al} using Chiral Perturbation Theory 
(ChPT) up to NNLO and reproducing low-energy $NN$ phase shifts with 
good precision \cite{prc.93.024002,prc.89.014002,prc.86.034005,prc.84.054009}.



The $N/D$ method \cite{chew.119.467} writes down a $NN$ partial wave as 
$T(A) = N(A)/D(A)$ 
  and such that  $N(A)$ and $D(A)$ have only LHC and right-hand cut (RHC), respectively.
 From elastic unitarity and the definition of $\Delta(A)$ as the LHC discontinuity of $T(A)$,
 $D(A)$ and $N(A)$ satisfy along their respective cuts:
\begin{eqnarray}
\label{030816.2}
{\rm Im}D(A)&=&-\rho(A)N(A)~,~A>0\\
{\rm Im}N(A)&=&\Delta(A) D(A)~,~A<L\nonumber
\end{eqnarray}
where $L=-m_\pi^2/4$ and $\rho(A)=M_N\sqrt A/4\pi$ 
 is the phase space, with $M_N$ the nucleon mass. 
In terms of the imaginary parts given in Eq.~\eqref{030816.2} one can write down in
 a standard way dispersion relations for the functions $D(A)$ and $N(A)$.
 The general form of these equations with an arbitrary number of subtractions can be found in Eq.(14) of Ref.~\cite{prc.89.014002}.
Here we use the $N/D$ method with the number of subtractions necessary to fit the effective range expansion
\begin{eqnarray}
  k\cot \delta = -\frac{1}{a} + \frac 1 2 r k^2 + \sum_{i=2} v_i k^{2i}
\end{eqnarray}
at a certain order. 
As in the $N/D$ method the functions $N(A)$ and $D(A)$ are defined up to a constant we have to perform at least one
subtraction, which is usually done for $D(A)$ fixing $D(0)=1$. The equations are:
\begin{align}
\label{021016.1}
D(A) =& 1 - i\frac{M_N \sqrt{A} }{4\pi^2}\int_{-\infty}^L d\omega_L \frac{D(\omega_L)\Delta(\omega_L)}{(\sqrt{\omega_L}+\sqrt{A})\sqrt{\omega_L}}
\nonumber
 \\
N(A) =& \frac{1}{\pi} \int_{-\infty}^L d\omega_L \frac{D(\omega_L)\Delta(\omega_L)}{(\omega_L-A)}
\end{align}
 We will call this approximation, which has no free parameters, 
 the regular case (or $N/D_{01}$), since for regular interactions the solutions are completely fixed by the potential  
 and must coincide with the ones given by the LS equation. 

If we want to fit the scattering length we can take an additional subtraction in $N(A)$ and we obtain the integral equation for the $N/D_{11}$ case, 
which reads
\begin{align}
  D(A) =& 1 + i a \sqrt A 
  + i\frac{A M_N}{4\pi^2} \int_{-\infty}^L d\omega_L 
  \frac{D(\omega_L)\Delta (\omega_L)}{\omega_L(\sqrt A+\sqrt{\omega_L})} \nonumber
\end{align}
where we see the independence on the subtraction point and the explicit dependence on the scattering length $a$. 

If we perform an additional subtraction in $D(A)$ we can fix also the effective range (case $N/D_{12}$).
Finally, by taking an additional subtraction in $N(A)$ we fix $v_2$ (case $N/D_{22}$). 
 Explicit expressions for these cases can be found in Refs.~\cite{prc.89.014002,forthcoming}.

It is worth mentioning here two points already discussed in Sec.II.A of Ref.~\cite{prc.93.024002}. 
 First, the number of subtractions in $N(A)$ is always less or equal to that taken 
in $D(A)$ because otherwise some of the RHC integrals in $D(A)$ would be divergent.
 Second,  any possible  Castillejo-Dalitz-Dyson pole \cite{pr.101.453} in the $D(A)$ function 
can be remove by taking one more subtraction at the same time  in $N(A)$ and $D(A)$. 

\begin{figure}[ht]
  \begin{center}
    \scalebox{1.0}{\includegraphics{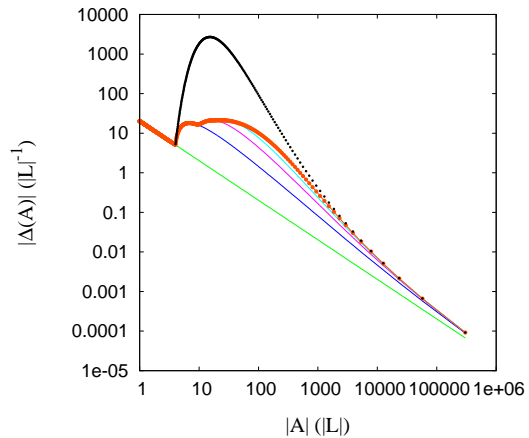}}
    \caption{\label{DeltaA1} Results for the $^1S_0$ LHC discontinuity $\Delta(A)$ for the
    unphysical value of $g_A=6.80$. 
    The solid lines shows $\Delta(A)$ including contributions
    of $1\pi$ (green), $2\pi$ (blue), $3\pi$ (magenta), $4\pi$ (light-blue) and the solution of Eq.~(\ref{Delta})
    (red points). The asymptotic expression Eq.~(\ref{DeltaAEc}) is shown by the black dots above $k>m_\pi$.}
  \end{center}
\end{figure}

One-pion exchange (OPE) at leading in  ChPT 
for the singlet $^1S_0$ partial wave can be written as
\begin{eqnarray}
\label{040816.1}
  V(p',p) &=& \frac{g_A^2}{4f_\pi^2} \frac{m_\pi^2}{2p'p} Q_0(z) + C_0
\end{eqnarray}
where $Q_0(z)=\frac{1}{2} \log[(z+1)/(z-1)]$ 
 and $z = (p'^2+p^2+m_\pi^2)/2p'p$.  
Here the delta-like contribution of the OPE potential has been included in the contact term $C_0$.
 The latter does not generate contribution to $\Delta(A)$ 
 (as well as any other polynomial that were added to the right-hand side of Eq.~\eqref{040816.1} corresponding to local terms). 
The contribution of OPE to $\Delta(A)$  stems entirely from the function $Q_0(z)$ evaluated on-shell, and is given by 
$\Delta_{1\pi}(A) = \theta(L-A) \frac{g_A^2}{4f_\pi^2} \frac{\pi m_\pi^2}{4A}$.
The once iterated OPE contribution using dimensional regularization has been evaluated by the
M\"unich group~\cite{npa.625.758} and its 
 contribution to the LHC discontinuity, already used within the $N/D$ method in Refs.~\cite{prc.93.024002,prc.89.014002}, is
\begin{widetext}
\begin{eqnarray}
\label{delta2pi}
  \Delta_{2\pi}(A) &=& \theta(4L-A)\left( \frac{g_A^2m_\pi^2}{16 f_\pi^2}\right)^2 \frac{M_N}{A\sqrt{-A}}\log\left(\frac{2\sqrt{-A}}{m_\pi}-1\right)
\end{eqnarray}
We have evaluated the contribution of twice and three-times 
 iterated OPE finding
\begin{eqnarray}
\label{delta4pi}
  \Delta_{3\pi}(A) &=& \theta(9L-A)\left(\frac{g_A^2m_\pi^2}{4f_\pi^2}\right)^3 \left(\frac{M_N}{4\pi}\right)^2 \frac{\pi}{4A}
	\int_{2m_\pi}^{2\sqrt{-A}-m_\pi} d\mu_1 \frac{1}{\mu_1(2\sqrt{-A}-\mu_1)}
  \nonumber \\ &&
	\theta(\mu_1-2m_\pi) \int_{m_\pi}^{\mu_1-m_\pi} d\mu_2 \frac{1}{\mu_2(2\sqrt{-A}-\mu_2)}
  \\
  \Delta_{4\pi}(A) &=& \theta(16L-A)\left(\frac{g_A^2m_\pi^2}{4f_\pi^2}\right)^4 \left(\frac{M_N}{4\pi}\right)^3 \frac{\pi}{4A}
  \int_{3m_\pi}^{2\sqrt{-A}-m_\pi} d\mu_1 \frac{1}{\mu_1(2\sqrt{-A}-\mu_1)}
  \nonumber \\ &&
  \theta(\mu_1-3m_\pi) \int_{2m_\pi}^{\mu_1-m_\pi} d\mu_2 \frac{1}{\mu_2(2\sqrt{-A}-\mu_2)}
  \theta(\mu_2-2m_\pi) \int_{m_\pi}^{\mu_2-m_\pi} d\mu_3 \frac{1}{\mu_3(2\sqrt{-A}-\mu_3)}.
\end{eqnarray}
For a diagram with $n$ pions we infer from the structure of the evaluated $\Delta_{m\pi}(A)$, $1\leq m\leq 4$, 
 that
\begin{eqnarray}
  \Delta_{n\pi}(A) &=& \theta(n^2L-A)\left(\frac{g_A^2m_\pi^2}{4f_\pi^2}\right)^n \left(\frac{M_N}{4\pi}\right)^{n-1} \frac{\pi}{4A}
  \prod_{j=1}^{n-1}
  \theta(\mu_{j-1}-(n+1-j)m_\pi) \int_{(n-j)m_\pi}^{\mu_{j-1}-m_\pi} d\mu_j \frac{1}{\mu_j(2\sqrt{-A}-\mu_j)}
\end{eqnarray}
with $\mu_0\equiv 2\sqrt{-A}$. This is the formal solution of the IE
\begin{eqnarray}
  \tilde \Delta(A,\bar \mu) &=& \Delta_{1\pi}(A) + \left(\frac{M_NA}{\pi^2}\right) 
  \theta(\bar \mu-2m_\pi)
  \int_{m_\pi}^{\bar \mu-m_\pi} d\mu 
  \frac{\Delta_{1\pi}(A)\tilde \Delta (A,\mu)}{\mu(2\sqrt{-A}-\mu)}
  \label{Delta}
\end{eqnarray}
\end{widetext}
such that $\Delta(A)=\tilde \Delta (A,2\sqrt{-A})$. 
 This IE gives the contribution to the LHC discontinuity 
in the $^1S_0$ partial wave of the iterated OPE contribution when solved for $\bar\mu\in[m_\pi,2\sqrt{-A}]$. 
 Notice that, the  denominator in the IE never vanishes and the limits of the integration are finite for $\sqrt{-A}<\infty$. 
A rigorous proof of Eq.~\eqref{Delta} and the generalization to other partial waves and interactions
will be given in a forthcoming paper \cite{forthcoming}. An important point to notice here is that in the general case, even for
singular interactions, $\Delta(A)$ is finite.  
 The divergent part of the LS equation in the LHC, if present, 
  would affect only 
the real part of $T$, which is not an input for the $N/D$ method.

When $\sqrt{-A}\gg m_\pi$ we can solve algebraically the IE of Eq.~\eqref{Delta}
 and obtain the asymptotic behavior of the LHC discontinuity which is given by
\begin{eqnarray}
  \label{DeltaAEc}
  \Delta(A) &=& \frac{\lambda\pi^2}{M_NA} e^{\frac{2\lambda}{\sqrt{-A}} {\rm arctanh} \left(1-\frac{m_\pi}{\sqrt{-A}}\right)}
\end{eqnarray}
with $\lambda = \frac{g_A^2 m_\pi^2}{4f_\pi^2} \frac{M_N}{4\pi}$.

For the physical case with $g_A=1.26$ the twice iterated OPE reproduces closely 
 the non-perturbative results.
 Because of this we consider the unphysical larger value\footnote{This value is chosen to reproduce 
 the physical scattering length $a=-23.75$~fm from the LS equation and the $N/D_{01}$ case. 
} $g_A=6.80$
 and  show in Fig.~\ref{DeltaA1} the comparison of the LHC discontinuity including
$\Delta_{1\pi}$ (green), $\Delta_{2\pi}$ (blue), $\Delta_{3\pi}$ (magenta), 
$\Delta_{4\pi}$ (light blue), the result of Eq.~(\ref{Delta}) (red points) and the asymptotic expression Eq.~(\ref{DeltaAEc})
(black points), which is only taken for $k>m_\pi$. In this case the contributions up to 4 pion exchanges are sizable.

\begin{figure}[ht]
  \begin{center}
    \scalebox{1.0}{\includegraphics{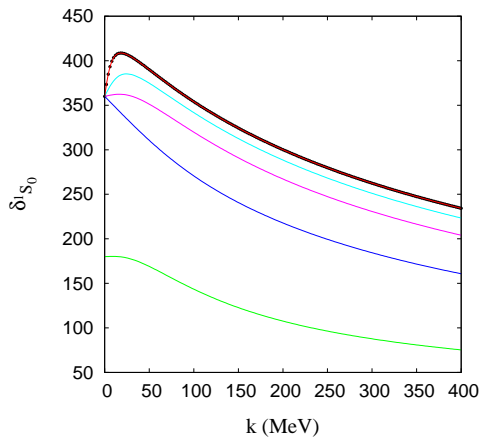}}
    \caption{\label{dPhysR01} Results for the $^1S_0$ phase-shift up to $k=400$ MeV with OPE for the N/D$_{01}$ case
    and $g_A=6.80$.
    The dots show the result of the LS equation while 
    the solid lines are the $N/D$ method including in $\Delta(A)$ contributions
    of $1\pi$ (green), $2\pi$ (blue), $3\pi$ (magenta), $4\pi$ (light-blue) and the solution of Eq.~(\ref{Delta})
    (red). 
}
  \end{center}
\end{figure}

Now that we have the contribution to the LHC discontinuity of the iterated OPE to any order, we can use the $N/D$ method to
 calculate the partial-wave amplitude along the physical region ($A>0)$. As OPE in the singlet case is a regular interaction
 we can perform a regular $N/D$ calculation and compare with the result of the LS equation. 
As mentioned before, OPE in the $^1S_0$ partial wave 
is weak, and
in order to see the effect of higher-order iterative diagrams in the calculation of $\Delta(A)$ 
we change the axial coupling constant to the unphysical value $g_A=6.80$. 
The results for the phase-shifts are given in Fig.~\ref{dPhysR01} where we can see that now the problem is non-perturbative;
 even including $4\pi$ exchange is not enough to reproduce the phase shift. However with the solution of Eq.~(\ref{Delta})
we reproduce the calculation of the LS equation. 
Note that here we take the prescription of zero
phase shift at infinite energy.\footnote{Levinson theorem, which is valid for regular interactions, implies
 that there are two bound states.}

\begin{center}
\begin{table}[ht]
\begin{tabular}{ccccccccc}
  \hline
  \hline
&  &  $a_s$ (fm)  &  $r$ (fm)    & $v_2$ (fm$^{3}$) \\ 
  \hline
  \hline
& $\Delta_{1\pi}$  &       -23.75 &         8.56 &        15.3  \\
& $\Delta_{2\pi}$  &       -23.75 &         8.80 &        17.7 \\
N/D$_{11}$ & $\Delta_{3\pi}$  &       -23.75 &         8.88 &        18.4  \\
& $\Delta_{4\pi}$  &       -23.75 &         8.90 &        18.6  \\
& Non-perturbative &       -23.75 &         8.90 &        18.7  \\
  \hline
  \hline
& $\Delta_{1\pi}$  &         1.66 &         0.714&       -0.168  \\ 
& $\Delta_{2\pi}$  &         3.53 &         2.03 & -5.70 $10^{-2}$  \\ 
N/D$_{01}$ & $\Delta_{3\pi}$  &         1.80 &         1.15 & -8.71 $10^{-2}$  \\ 
& $\Delta_{4\pi}$  &        -6.89 &         13.7 &         47.5  \\ 
& Non Perturbative &       -23.75 &         8.90 &        18.7  \\ 
  \hline
  \hline
\end{tabular}
\caption{\label{EfectrR} Effective range parameters for the $N/D$ calculation 
for the unphysical value of $g_A=6.80$.}
\end{table}
\end{center}

\begin{figure}[ht]
  \begin{center}
    \scalebox{1.0}{\includegraphics{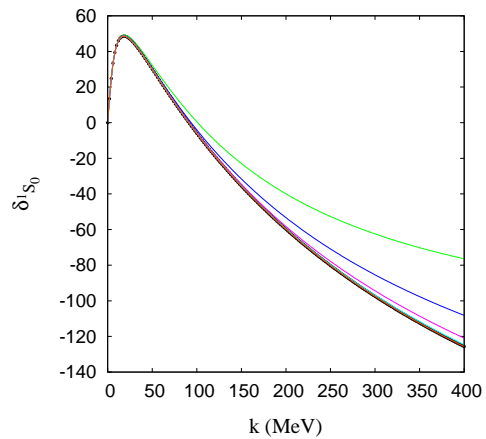}}
    \caption{\label{dPhysR11} Results for the $^1S_0$ phase-shift up to $k=400$ MeV with OPE for the $N/D_{11}$ case
    and $g_A=6.80$.
    Lines are the same as in Fig.~\ref{dPhysR01}.}
  \end{center}
\end{figure}

At this point it is interesting to check what we get if use the $N/D$ method with one subtraction fixing the scattering
length to the value obtained from the LS equation. This is given in Fig~\ref{dPhysR11} where we can see
that if we include in $\Delta(A)$ the solution of Eq.~(\ref{Delta}) we recover exactly the result without subtractions.
However in this case the contributions of the iterative diagrams to $\Delta(A)$ look perturbative and with $4\pi$ exchange
we almost recover the exact result.
Note that here we take the prescription of zero
phase shift at threshold.

In Table~\ref{EfectrR} we give the  first three 
effective range parameters  for both calculations, the regular one and 
the $N/D_{11}$. For the latter the parameters vary slowly  when more pion exchanges are added. 
For the regular solution, due to the fact that the system is highly non-perturbative,
we do not see a convergent pattern, it will probably show up at a much higher order. However the final result
in both prescriptions agrees.


These considerations show how a non-perturbative problem (with respect to the contributions from the LHC discontinuity to the 
physical partial-wave amplitude)   is almost perturbative once one extra 
subtraction (or more) are taken.

We now compare the results given by the $N/D$ method with those of the LS
equation renormalized with one counter term. We use three different regulators. The first one
corresponds to a monopole form factor which gives the partial wave projection
\begin{eqnarray}
\label{070816.1}
  V_1(p',p) &=& C_0^1 \frac{\Lambda^2}{2p'p} Q_0(z_\Lambda)
\end{eqnarray}
where $z_\Lambda = \frac{p'^2+p^2+\Lambda^2}{2p'p}$
and has the same analytical properties as the OPE.
The second and third ones are Gaussian type form factors
\begin{eqnarray}
  V_i(p',p) &=& C_0^i e^{-\left(\frac{p'}{\Lambda}\right)^{2n_i}-\left(\frac{p}{\Lambda}\right)^{2n_i}}
\end{eqnarray}
where $n_2=2$ and $n_3=3$ and have
different analytical properties. In all cases for each value of $\Lambda$
 we determine $C_0^i$ to reproduce the $^1S_0$ scattering length $a=-23.75$~fm. 
 We also calculate the LS equation  using subtractive renormalization~\cite{npa.653.209,prc.80.044002}.
 It has the advantage that the
 counter term is removed from the equations and they depend explicitly on the low energy constant fixed, e.g. the scattering length $a$.

\begin{figure}[t]
  \begin{center}
    \scalebox{1.0}{\includegraphics{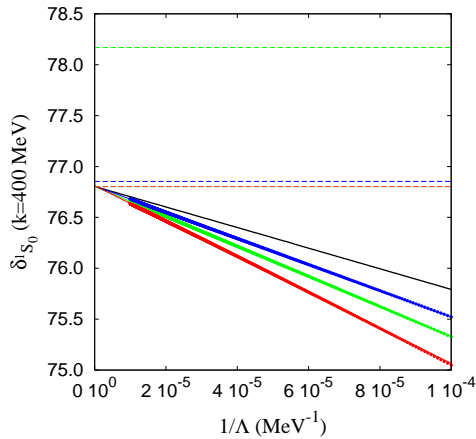}}
    \caption{\label{d400} Results for the $^1S_0$ phase-shift at $k=400$ MeV with OPE as a function of
    $1/\Lambda$. The dots show the result of the LS renormalized 
 with one counter term as in $V_1$ (red), $V_2$ (green) and $V_3$ (blue).
    A linear fit to extrapolate to $\Lambda\to \infty$ is also shown.
    The solid black line gives the result from subtractive renormalization.
    The dashed lines show results of the $N/D$ method with one extra subtraction including in $\Delta(A)$ contributions
    of $1\pi$ (green), $2\pi$ (blue), $3\pi$ (magenta), $4\pi$ (light-blue) and the non-perturbative
    calculation (red). 
}
  \end{center}
\end{figure}

To compare with the result of the $N/D$ method we have to make an extrapolation to $\Lambda\to \infty$. 
 In order to do that we plot in Fig.~\ref{d400} the phase shift for $k=400$ MeV as a function of $1/\Lambda$. 
In the Figure the dots represent the phase-shifts obtain with the LS equation with the counter terms
introduced 
  as in $V_1$ (red), $V_2$ (green) and $V_3$ (blue).
These plots show a linear behavior and we make a linear fit to obtain the phase-shift at $1/\Lambda=0$.
The solid black line shows the result from subtractive renormalization~\cite{npa.653.209,prc.80.044002}.
The dashed lines correspond to the results obtained from the case $N/D_{11}$  with the same
colors as in Fig.~\ref{dPhysR01}. In this scale contributions
from $3\pi$ exchange can be observed. The final result agrees very well with the extrapolation made
by the fit and this does not depend on the particular regulator used.


\begin{figure}[t]
  \begin{center}
    \scalebox{1.0}{\includegraphics{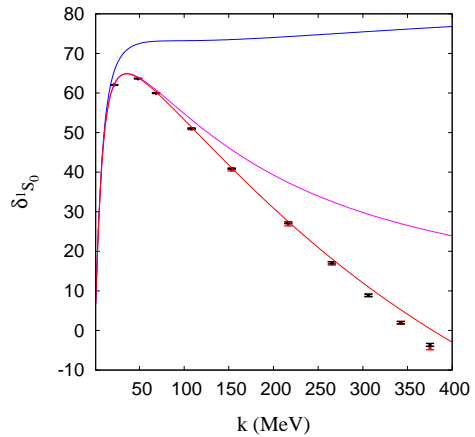}}
    \caption{\label{LO} The $^1S_0$ phase-shift obtained from OPE and the $N/D$ method. 
 The black dots with errorbars are
    the Granada  $np$  phase-shift analysis~\cite{prc.88.064002} and the red dots with errorbars are the
    Nijmegen 93 partial wave analysis~\cite{prc.48.792}. The solid blue line is the result
    with one extra subtraction, the magenta with two and the red with three.}
  \end{center}
\end{figure}

Finally, we give in Fig.~\ref{LO} the phase shifts that result in the $N/D$ method with $\Delta(A)$ calculated from
 the LO ChPT potential. The cases $N/D_{11}$, $N/D_{12}$ and $N/D_{22}$ are shown. 
 The subtraction constants are fixed  using
 $a=-23.75$ fm, $r=2.66$ fm and $v_2=-0.63$~fm$^3$.

We now perform the $N/D$ method calculation with singular interactions given by the NLO and NNLO Chiral
Effective Field Theory contributions~\cite{npa.625.758}. At NNLO we use the
LEC's $c_1=-0.74$ GeV$^{-1}$, $c_3=-3.61$ GeV$^{-1}$ and $c_4=2.44$ GeV$^{-1}$ obtained
from the $\pi N$ Roy-Steiner equations matched to ChPT~\cite{prl.115.192301}.
 However, due to the singular character of the interaction not all the $N/D$ equations converge 
 (contrary to the regular case of the Yukawa potential at LO). 
Indeed, 
since the interactions are singularly attractive one would expect no solution for the regular case and a well defined solution
for the $N/D_{11}$ case, and this is what we obtain.\footnote{This expectation is based on the fact that 
a singularly attractive interaction requires to fix a constant relative phase \cite{pr.80.797,prc.77.044006}.}
 The case $N/D_{12}$ does not converge, which explains from first principles why when trying to fix  
only $a$ and $r$  no solutions were obtained in the LS studies of  Refs.~\cite{prc.77.044006,prc.80.044002}. The 
  $N/D_{22}$ equations give a well defined result. 
Detailed discussions on the calculation of $\Delta(A)$
  and about the convergence of different calculations will be given in a forthcoming paper \cite{forthcoming}.

\begin{figure}[t]
  \begin{center}
    \scalebox{1.0}{\includegraphics{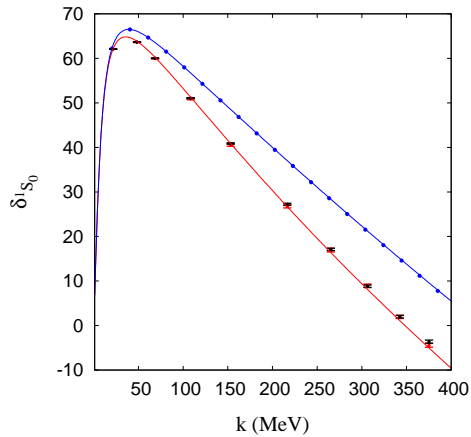}}
    \caption{\label{NLO} The $^1S_0$ phase-shift  from the NLO ChPT potential. 
 The solid blue dots are the result of the LS equation with subtractive renormalization. 
The meaning of the dots is the same as in Fig.~\ref{LO} and each line corresponds to the same type of 
$N/D$ equations as there.
}
  \end{center}
\end{figure}

\begin{figure}[t]
  \begin{center}
    \scalebox{1.0}{\includegraphics{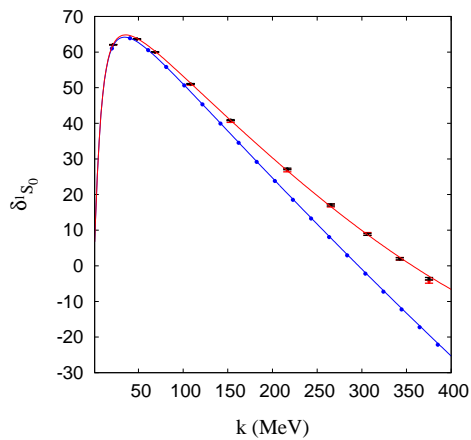}}
    \caption{\label{NNLO} The $^1S_0$ phase-shift from the NNLO ChPT potential. For the meaning of the 
lines and dots see Fig.~\ref{NLO}. 
}
  \end{center}
\end{figure}

In Figs.~\ref{NLO} and~\ref{NNLO} we show the results that stem from the calculations employing the 
NLO and NNLO ChPT potentials, respectively. We use the same values for $a$, $r$ and $v_2$ and the same
colors as in Fig.~\ref{LO}, additionally 
we also include as blue dots the result of subtractive renormalization.

In this work we did not perform a thorough fit of the phase shifts since the electromagnetic and
charge symmetry breaking effects, which are important at low energies, has not been included
yet. However, this work shows that the $^1S_0$ partial wave could be reproduced with good 
accuracy with the NNLO ChPT potential 
 within the $N/D_{22}$-case calculation.

A remarkably interesting feature of the $N/D$ method over the LS equation
 is that one can calculate straightforwardly the scattering amplitude in the whole $A$-complex plane, once $D(A)$ 
is known along the LHC, cf. Eq.~\eqref{021016.1}. In this way, 
one could study directly the bound states, virtual states and resonances associated with the calculated partial-wave amplitude.
 The large scattering length in the $^1 S_0$ partial wave is a signal of an antibound state. We can calculate the
$T$ matrix in the second Riemann sheet [$T_{II}(A)=N(A)/D_{II}(A)$] using the analytical continuation of the first Riemann sheet given by
\begin{eqnarray}
	T_{II}^{-1}(A) &=& T^{-1}(A) + 2i \rho(A) 
\end{eqnarray}
where $\rho(A)$ is evaluated such that  ${\rm Im}(\sqrt A)>0$. So
the antibound state is given by the zeros of $D_{II}(A)$. 
With one subtraction  we obtain $-0.070$~MeV at LO, and $-0.067$~MeV
 for the NLO and NNLO cases. Using more than 
one subtraction we get $-0.066$ MeV in all cases.


 In summary, we have deduced for the first time in the literature an IE to obtain the full discontinuity 
of a partial-wave amplitude along the LHC, $\Delta(A)$. This is then implemented within the $N/D$ method, with 
exact discontinuities both along the LHC and RHC. In this way, we obtain  the equivalence between the $N/D$ method 
 and the the LS equation for a Yukawa potential (regular potential).  
We also apply the method with the  NLO and NNLO ChPT potentials, which are  examples of singular attractive interactions \cite{prc.77.044006}. 
 The equivalence of the  $N/D$ method with one extra subtraction  
and the LS equation renormalized with one counter term or with subtractive renormalization  holds in this case as well.
Within the $N/D$ method one can also include extra subtractions constants; we have done it up to three extra subtractions,
 which reproduces accurately the $^1S_0$ $NN$ phase shifts of the Granada group analysis~\cite{prc.88.064002} when  
 the NNLO ChPT potential is used. This goes definitely beyond the present 
nonperturbative solution of the LS equation in momentum space renormalized with counter terms, 
for which theoretical control in the regulator independent case  
is achieved only when taking  one or none counter term \cite{prc.77.044006,prc.80.044002}. 
Our results are far reaching and 
could be of great interest for atomic, molecular, nuclear and particle physics in which singular attractive potentials and
short range interactions usually appear.


\acknowledgments
This work has been partially funded by MINECO
under Contract No. FPA2013-47443-C2-2-P, by the MINECO (Spain) and 
ERDF (European Commission) grant FPA2013-40483-P and by
the Spanish Excellence Network on Hadronic Physics FIS2014-57026-REDT.

\bibliographystyle{apsrev}
\bibliography{NDletter.bib}

\end{document}